\documentclass[12pt,preprint]{aastex}

\newcommand{\mdot}{M$_{\odot}$ yr$^{-1}$}
\newcommand{\ldot}{L$_{\odot}$}
\newcommand{\um}{$\mu$m~}
\newcommand{\ums}{$\mu$m}

\def\kmsMpc{\ifmmode {\rm\,km\,s^{-1}\,Mpc^{-1}}\else
    ${\rm\,km\,s^{-1}\,Mpc^{-1}}$\fi}

\begin{document}

\title{From IRAS to IRS: Evolution of the Most Luminous Galaxies in the Universe}
\author{James R. Houck and Daniel W. Weedman}
\affil{Astronomy Department, Cornell University, Ithaca, NY 14853, jrh13@cornell.edu, dweedman@isc.astro.cornell.edu}
\begin{abstract}
  We summarize observations with the $Spitzer$ Infrared Spectrograph (IRS) of 571 starbursts (strong PAH emission features), 128 obscured AGN (strong silicate absorption), and 39 unobscured AGN (silicate emission).  Sources range in luminosity from 10$^{8}$ to 10$^{14}$ \ldot~ and continuously in redshift for 0 $<$ z $<$ 3.  The most luminous starbursts and AGN evolve as (1+z)$^{2.5}$ to z $\sim$ 2.5; no clear evidence is found that this evolution ceases beyond z = 2.5.  Dust obscuration in starbursts is determined by comparing PAH luminosity with ultraviolet luminosity and indicates severe obscuration in most starbursts, even those selected in the ultraviolet; the median ratio (intrinsic ultraviolet/observed ultraviolet) is $\sim$ 50 for infrared selected starbursts and $\sim$ 8 for ultraviolet selected starbursts.  Obscuration increases with bolometric luminosity, but starbursts which appear most luminous in the ultraviolet are those with the least obscuration.  This result indicates that extinction corrections are significantly underestimated for ultraviolet selected sources, suggesting that galaxies at z $>$ 2 are more luminous than deduced only from rest frame ultraviolet observations. 
\end{abstract}
\section{Introduction}
At the same time that the $Spitzer$ mission was being conceived as NASA's final Great Observatory (originally as SIRTF), results from the Infrared Astronomical Satellite (IRAS) demonstrated that the most luminous galaxies in the universe are dusty and heavily obscured, with most of their luminosity arising from optical and ultraviolet luminosity which is reradiated by the absorbing dust \citep*{soi87}.  A fundamental motive of $Spitzer$ was to track the evolution of these luminous galaxies through the history of the Universe.  In what follows, we review the success of $Spitzer$ in achieving this goal.


Our conceptual sketch of such an ultraluminous infrared galaxy (ULIRG) is shown in Figure 1 and described in the caption.  Undoubtedly, starbursts are present to some extent in all such sources.  The starburst ULIRGs are the most fertile star forming regions in the Universe.  However, the observed characteristics are determined both by the luminosity of the starburst and the luminosity of dust heated by a central active galactic nucleus (AGN).  Here, we separate the two phenomena to track separately the evolution of starbursts and AGN.

The plots and results that we show below derive only from IRS low resolution observations.  More detailed diagnostics can be made for those sources with high resolution data \citep[e.g.][]{far07,ber09}, but the heavy extinction in the most obscured sources can hide even infrared emission lines arising near the central AGN.  For example, the presence of only blueward asymmetries on infrared forbidden lines from heavily obscured ULIIRG AGN indicates that the obscuring dust is thick, even at the scale of the narrow line region of AGN \citep{spo09}.  
\begin{figure}[!ht]
\figurenum{1}
\includegraphics[scale=0.25]{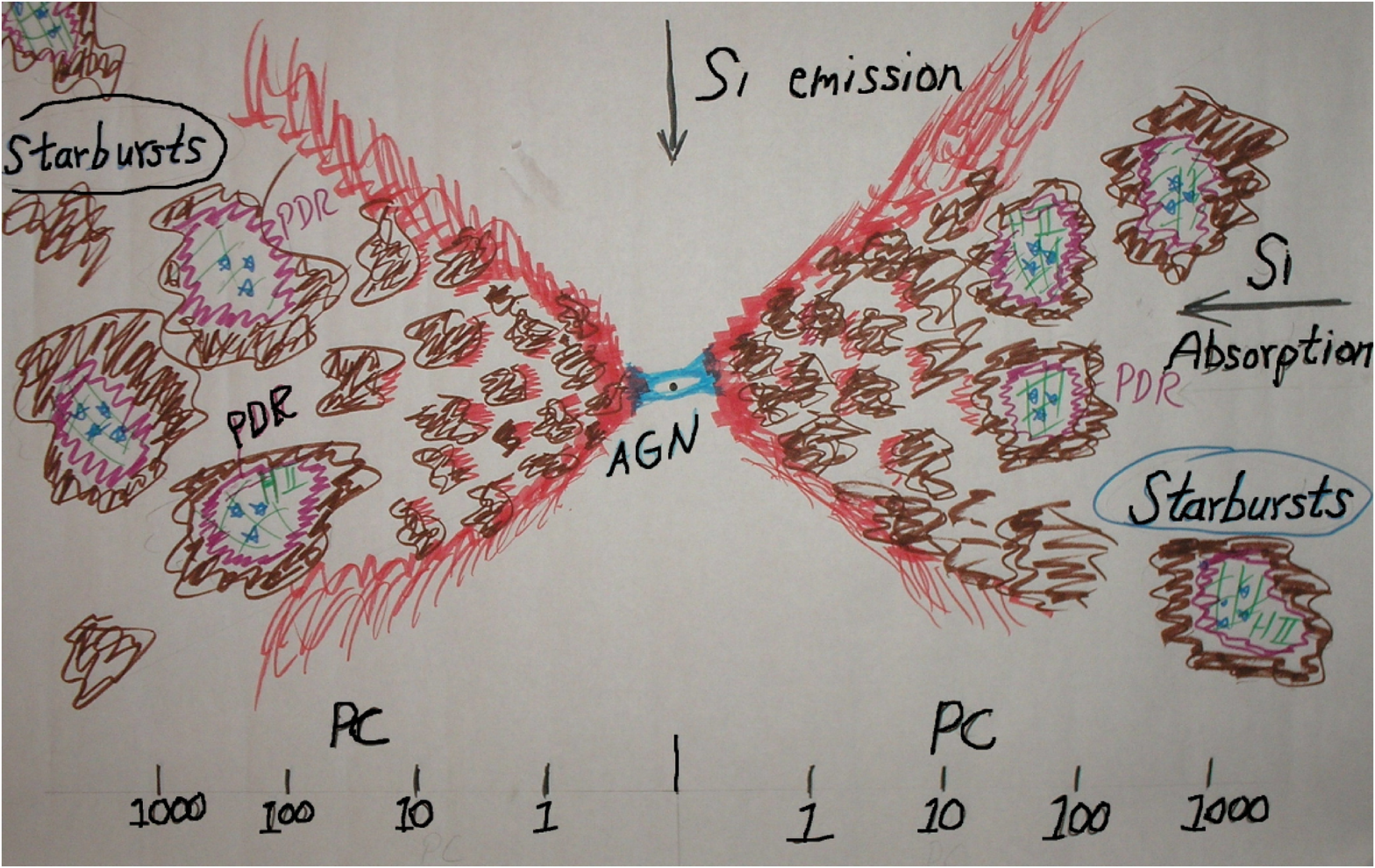}
\caption{Sketch of the assumed model for a luminous, dusty galaxy.  Starbursts can be found within the large torus containing molecular clouds and dust; each starburst immersed in a molecular cloud is surrounded by a photodissociation region (PDR) which produces strong PAH emission features.  Any active galactic nucleus (AGN) heats dust close to the nucleus.  (Note logarithmic size scale at bottom.) The starbursts will appear the same from any viewing direction, but the appearance of the AGN depends on whether it is viewed face on or through the torus.  If face on, the hottest dust is seen directly and silicate emission is seen.  If seen through the torus, the dust continuum from the hot dust is absorbed within cooler dust and produces silicate absorption.  The final observed spectrum is some combination of PAH, silicate absorption, or silicate emission depending on the relative luminosities of starbursts and AGN, and on the viewing direction for the AGN. }
\end{figure}
In general, the new IRS results confirm the initial demonstration by \citet{gen98} that strong emission from polyclyclic aromatic hydrocarbons (PAH) is associated with starbursts, but not with AGN.  The photodissociation region at the interface between the surrounding molecular cloud and the HII region of a starburst provides the appropriate physical conditions for strong PAH excitation and emission \citep{pee04}.  The hundreds of low resolution spectra accumulated by the IRS give an empirical result that "pure" starbursts - sources with no evidence in any spectral region of an AGN - always have equivalent width for the 6.2 \um PAH feature, EW(6.2 \ums), to be EW(6.2 \um)$>$ 0.4 \um \citep*{bra06,hou07,wee09a,sar10}.  

Conversely to starbursts, there are large numbers of sources with IRS spectra which are "pure" AGN, having no evidence in any spectral region of an observationally significant starburst; these have EW(6.2 \ums)$<$ 0.1 \ums.  We have used these simple observational criteria to derive the samples of starbursts and AGN presented below, as explained in \citet{hou07} and \citet{wee08,wee09a,wee09b}.  
\begin{figure}[!ht]

\figurenum{2}
\includegraphics[scale=0.95]{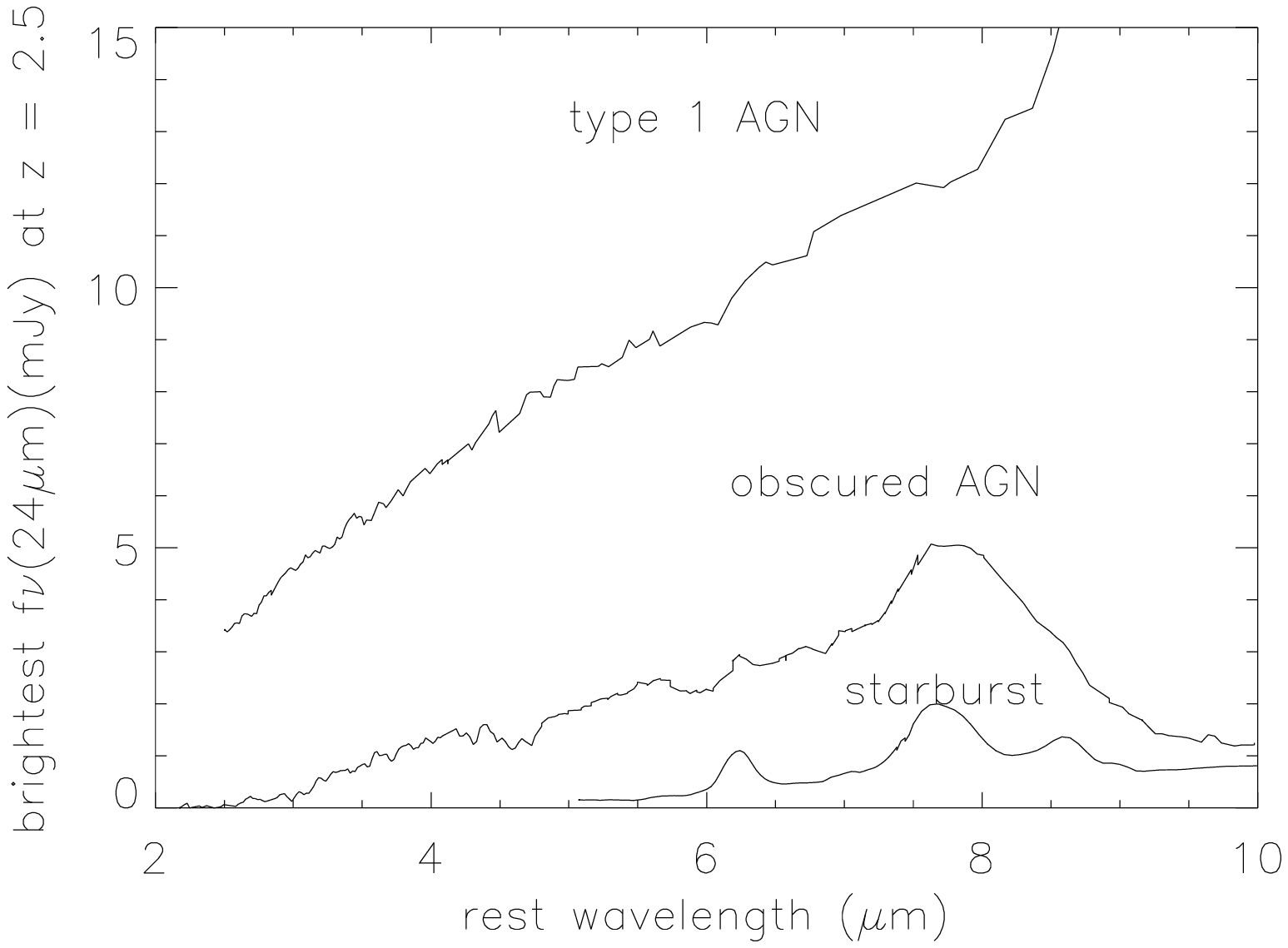}
\caption{Representative rest frame spectra observed for the most luminous dusty galaxies at z = 2.5.  The top spectrum represents the brightest type 1 AGN and is featureless at these rest wavelengths, but silicate emission begins to increase at the longest wavelengths toward a peak at $\sim$ 10 \ums.  Such sources are optically bright and have known redshifts, so IRS spectral measures of the dust continuum are made at 7.9 \ums.  The center spectrum is a heavily obscured AGN with an apparent spectral peak at 7.9 \um arising because of deep silicate absorption at longer wavelengths and other absorption features at shorter wavelengths.  The dust continuum is measured at this 7.9 \um peak.  The lowest spectrum is a starburst with strong PAH features.  Starburst luminosity is measured by the peak of the strong 7.7 \um PAH feature.  (The illustrated spectra have higher S/N than individual sources actually observed at such redshifts; spectra which are shown derive from brighter, lower redshift examples.)}
\end{figure}
 Our primary motivation for using the simple observational measurement of peak flux densities for the PAH 7.7 \um feature in spectra of "pure" starbursts or the 7.9 \um spectral peak for "pure" obscured AGN is to have reliable and self-consistent data for the large numbers of sources discovered by $Spitzer$ at z $\ga$ 2, for which S/N is poor, and deconvolution of the total flux within spectral features is uncertain.   

To measure total infrared luminosities, we have used as an initial calibration a limited number of ULIRGs or starbursts having IRAS fluxes and IRS spectra, so that L$_{bol}$ can be estimated as in \citet{san96}. These calibrations, from \citet{wee09b}, are given in the captions of Figures 3 and 4.  This estimate of L$_{bol}$ arises only from the IRAS wavelengths, with an assumed correction for colder dust components, so our calibration will be improved as more far infrared and submillimeter observations accumulate.  We use this L$_{bol}$ to calibrate the star formation rate (SFR) according to the precepts of \citet{ken98}, as given in the caption to Figure 3. 
\begin{figure}[!ht]

\figurenum{3}
\includegraphics[scale=0.95]{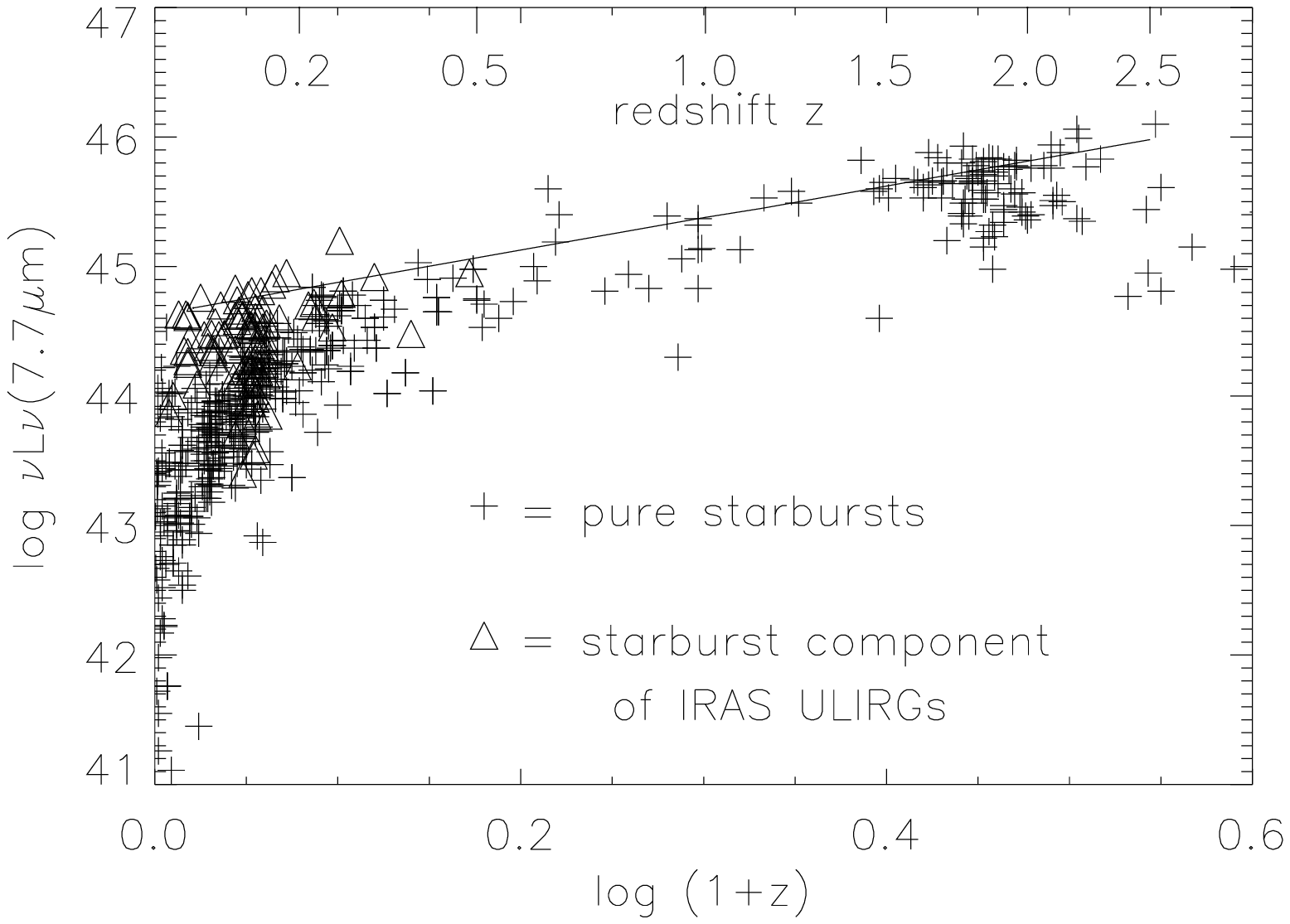}
\caption{ Starbursts with IRS measures of PAH luminosity, defined as $\nu$L$_{\nu}$(7.7$\mu$m) for the peak f$_{\nu}$(7.7$\mu$m), in units of ergs s$^{-1}$.  Line is the formal fit to evolution of the most luminous sources over all redshifts, which has form (1+z)$^{2.5}$.  Using our adopted calibration, log L$_{bol}$ = log $\nu$L$_{\nu}$(7.7$\mu$m) - 32.80 for L$_{bol}$ in \ldot~and $\nu$L$_{\nu}$(7.7$\mu$m) in ergs s$^{-1}$;  this yields that log(SFR) = log$\nu$L$_{\nu}$(7.7$\mu$m) - 42.57 for SFR in \mdot.} 
\end{figure}
All sources shown below were observed with the IRS, described in \citet{hou04}, using only the low resolution modules, giving spectral
coverage from $\sim$5\,\um to $\sim$35\,\ums.  For all spectra which we analyzed, spectral extractions were done with the SMART analysis package \citep{hig04}, usually with the improved "optimal extraction" procedure in \citet{leb10}\footnote{http://isc.astro.cornell.edu/IRS/SmartRelease}.
\section{Evolution of the Most Luminous Starbursts and AGN }
The spectral features at rest frame wavelengths $\sim$ 8 \um which we use to make uniform measures of source luminosity can be traced with the IRS continuously with redshift for 0 $<$ z $\la$ 3.  In Figures 3 and 4, we show the results for 571 starbursts and 167 AGN measured with the IRS.  Results for starbursts are taken from the measurements and summary in \citet{wee08} which includes published results from \citet{bra06}, \citet{hou07}, \citet{brn08b}, \citet{wee06a}, \citet{far08}, \citet{hou05}, \citet{wee06b}, \citet{yan07}, \citet{pop08}, \citet{far07}, \citet{ima07}, and \citet{sar08}; additional measurements from \citet{sar09} and \citet{sar10}; and additional published results from \citet{des09}, \citet{das09}, \citet{hua09}, and \citet{men09}.  
\begin{figure}[!ht]

\figurenum{4}
\includegraphics[scale=0.95]{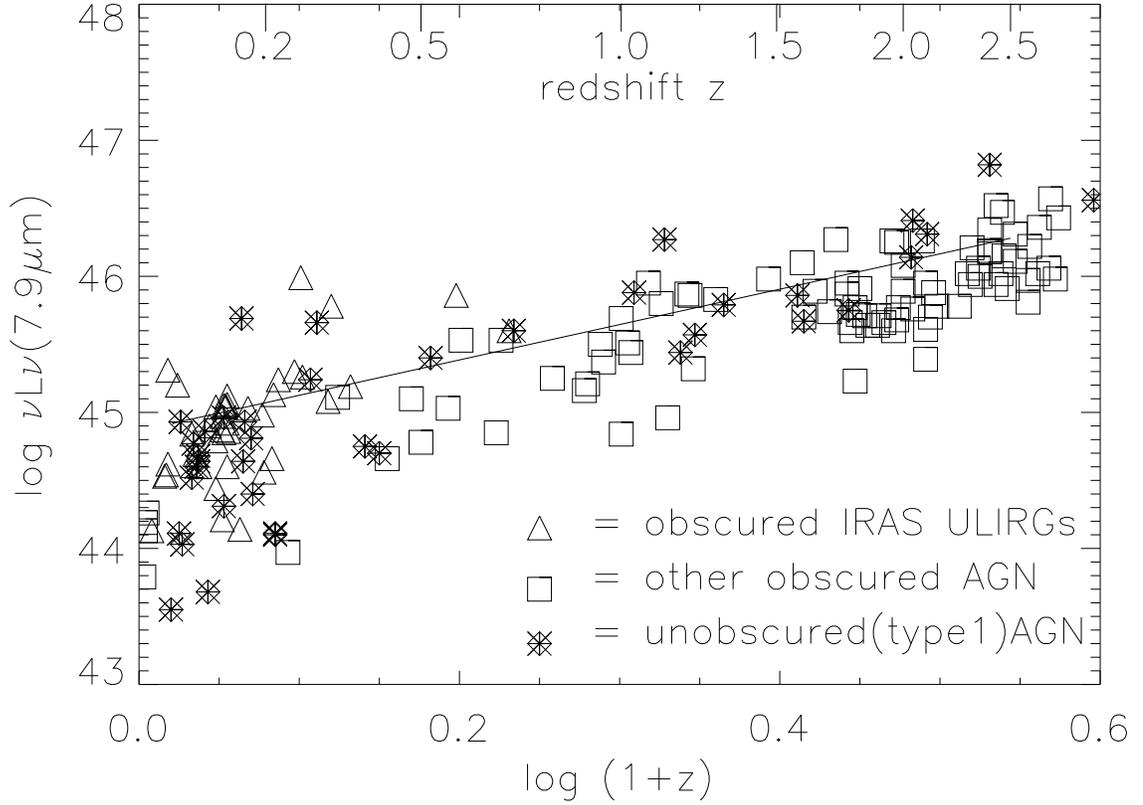}
\caption{All luminous AGN with dust continuum luminosity measured as $\nu$L$_{\nu}$(7.9$\mu$m), in units of ergs s$^{-1}$.  Obscured AGN are those having deep silicate absorption such that the 10 \um silicate feature absorbs more than 50\% of the extrapolated unabsorbed continuum at 10 \ums; unobscured AGN have silicate emission. Line is the formal fit to evolution of the most luminous obscured AGN over all redshifts, which has form (1+z)$^{2.6}$.  Using our adopted calibration for obscured AGN, log L$_{bol}$ = log $\nu$L$_{\nu}$(7.9$\mu$m) - 32.63, for L$_{bol}$ in \ldot~and $\nu$L$_{\nu}$(7.9$\mu$m) in ergs s$^{-1}$; for unobscured AGN, the adopted calibration is log L$_{bol}$ = log $\nu$L$_{\nu}$(7.9$\mu$m) - 32.84, indicating that bolometric luminosities are similar for obscured and unobscured AGN. } 
\end{figure}
Results for AGN are taken from the measurements and summary in \citet{wee09b} which includes published results from \citet{far07}, \citet{arm07}, \citet{ima07}, \citet{sar08}, \citet{shi07}, \citet{wee09a}, \citet{brn08a}, \citet{far09}, \citet{brn08b}, \citet{hou05}, \citet{wee06a}, \citet{wee06b}, \citet{pol08}, \citet{yan07}, \citet{saj08}, \citet{mar08}, \citet{hao05}, \citet{sch07}, and \citet{mk07}; and additional published results in \citet{her09}. 

The most important results in Figures 3 and 4 are that the form of luminosity evolution is very similar for both starbursts and AGN, and that no definitive indication is found for the epoch at which luminosity evolution reaches a maximum.  These results mean that whatever process triggers the formation of luminous, dusty galaxies controls the evolution of both starbursts and AGN, but we cannot yet determine whether starbursts or AGN came first.
\begin{figure}[!ht]

\figurenum{5}
\includegraphics[scale=0.95]{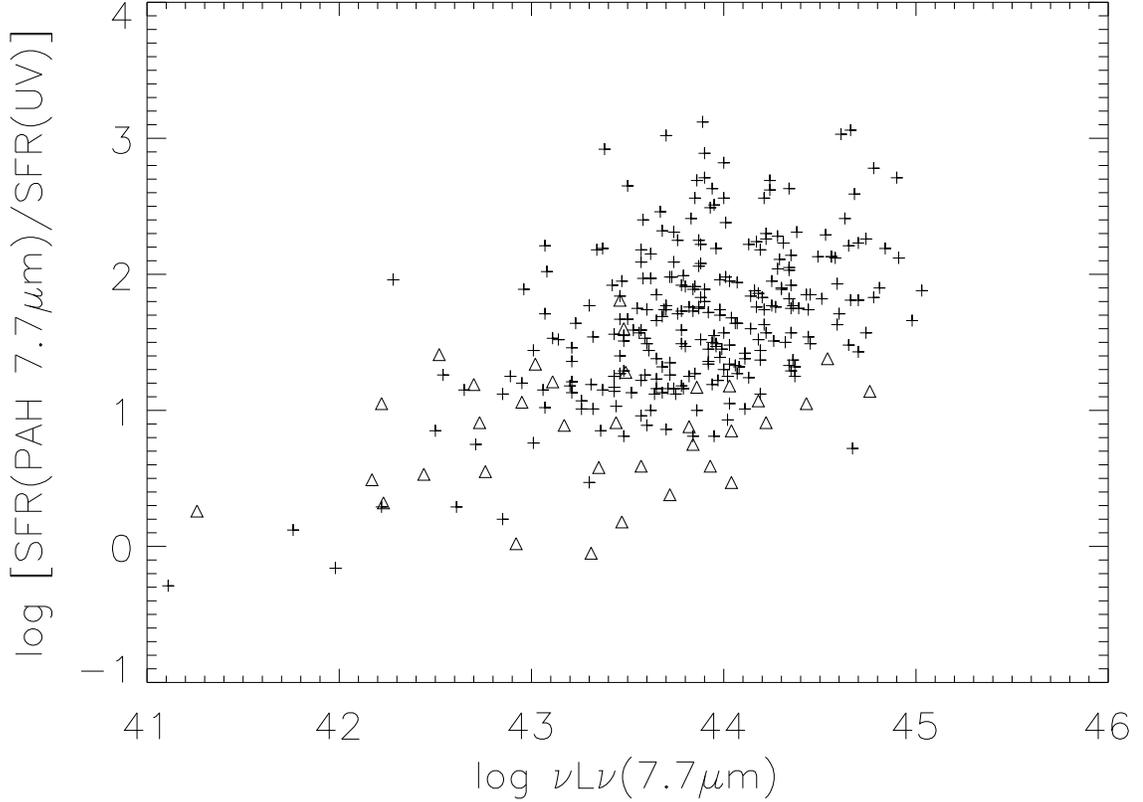}
\caption{Comparison of star formation rates measured from infrared PAH luminosity compared to observed ultraviolet luminosity; crosses are starbursts selected from $Spitzer$ MIPS 24 \um surveys and triangles are starbursts selected in the ultraviolet, either as luminous GALEX sources or as Markarian galaxies.  Luminosity $\nu$L$_{\nu}$(7.7$\mu$m) is in units of ergs s$^{-1}$ as in Figures 3 and 6. Using our adopted calibration, log L$_{bol}$ = log $\nu$L$_{\nu}$(7.7$\mu$m) - 32.80 for L$_{bol}$ in \ldot~and $\nu$L$_{\nu}$(7.7$\mu$m) in erg s$^{-1}$, and log SFR(7.7\um) = log $\nu$L$_{\nu}$(7.7$\mu$m) - 42.57 for SFR in \mdot.  Ratio of deduced SFRs indicates large dust extinction of the intrinsic ultraviolet luminosity, shown in Figure 6.} 
\end{figure}
\section{Star Formation Rates and Obscuration of Starbursts}
Tracing the evolution of star formation in the universe is a fundamental objective of observational cosmology.  The star formation rate (SFR) can be estimated with many observational techniques \citep[e.g.][]{ken98,cal08}.  Starbursts have been observed to the highest redshifts (z $\ga$ 7) using rest frame ultraviolet luminosities \citep[e.g.][]{bou09}, but determination of intrinsic source luminosities and true SFRs requires large corrections for obscuration by dust.  The availability of infrared spectra showing the PAH features, suffering much less extinction than optical or ultraviolet features, allows a new test of obscuration corrections \citep{sar09,sar10}.   Results are shown in Figures 5 and 6.  
Figure 5 illustrates the dramatic contrast in SFRs that would be measured from the PAH 7.7 \um feature, compared to SFRs measured from the ultraviolet continuum (details in \citet{sar09}). Even for starbursts selected in the ultraviolet, the differences are large.  This implies large extinction for the observed ultraviolet, quantified in Figure 6.
\begin{figure}[!ht]

\figurenum{6}
\includegraphics[scale=0.95]{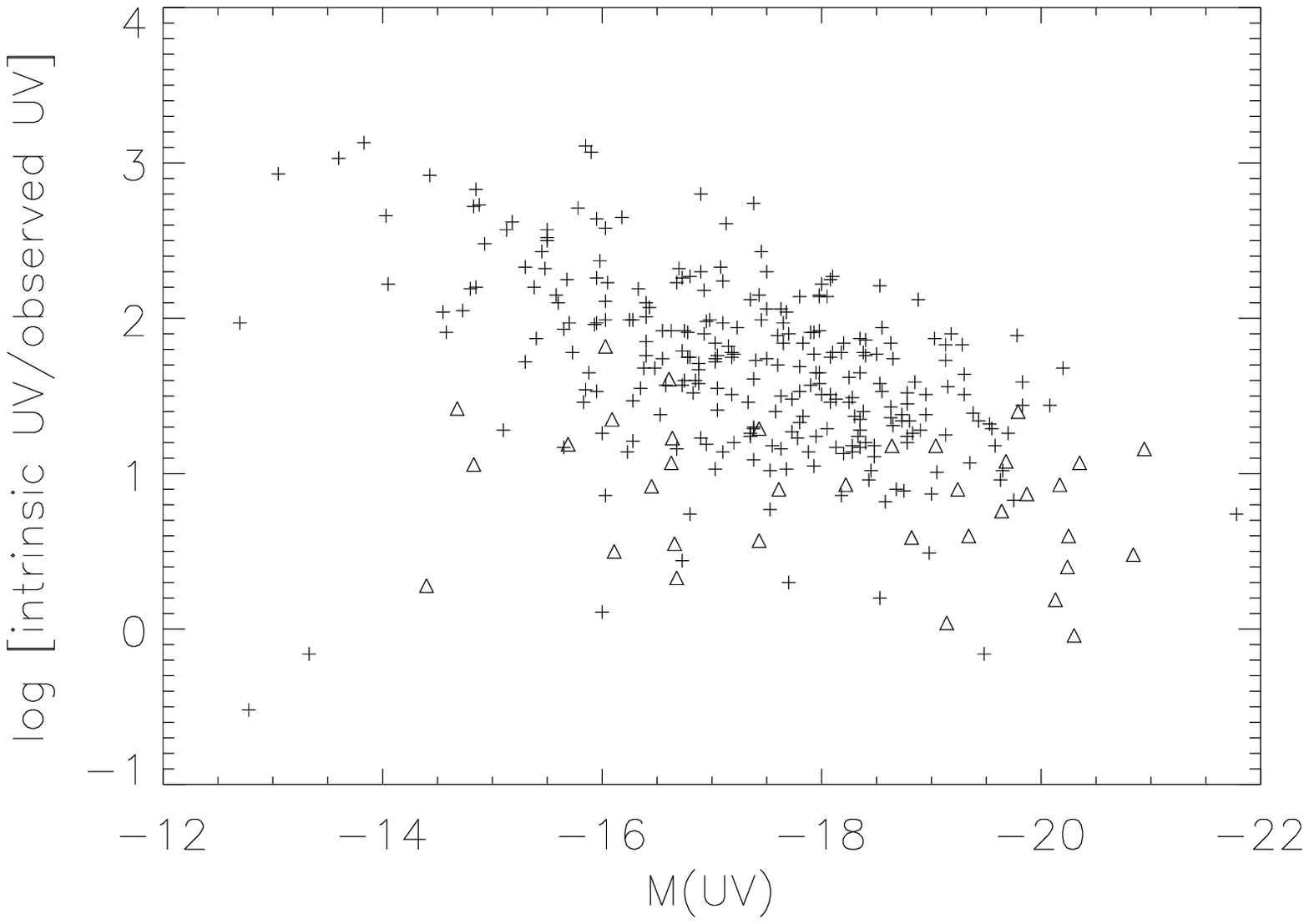}
\caption{Extinction factor for intrinsic ultraviolet luminosity, as determined from observed ratio f$_{\nu}$(7.7 $\mu$m)/f$_{\nu}$(153 nm).  Crosses are starbursts selected from $Spitzer$ MIPS 24 \um surveys and triangles are starbursts selected in the ultraviolet, either as GALEX sources or as Markarian galaxies.} 
\end{figure}
Using the smallest observed values of f$_{\nu}$(7.7 $\mu$m)/f$_{\nu}$(153 nm) to determine empirically the value of this ratio for unobscured starbursts, \citet{sar10} measure the amount by which the ultraviolet luminosity is attenuated by obscuring dust.  Figure 6 shows that the extinction is much greater in many infrared selected starbursts, but extinction is still high even in ultraviolet selected starbursts.  Required corrections for dust obscuration are higher by factors of 2 to 3 than generally assumed for the ultraviolet selected starbursts used to track luminosity evolution for z $\ga$ 2 \citep{red09}.  It is crucial to confirm whether these large corrections for obscuration are necessary at high redshift. 
\acknowledgments
We thank Don Barry for technical assistance, Lusine Sargsyan for assistance in data analysis, and V. Lebouteiller, Jeronimo Bernard-Salas and G. Sloan for developing the SMART optimal extraction software.  Support for this work by the IRS GTO team at Cornell University was provided by NASA through Contract 1257184 issued by JPL/Caltech.

\end{document}